\documentclass[aps,prb,twocolumn,showpacs,superscriptaddress,floatfix,citeautoscript,floatfix]{revtex4-1}
\pdfoutput=1
\usepackage{dcolumn}
\usepackage{bm}
\usepackage{amsmath, amscd, amssymb, amsthm}
\usepackage{times}
\usepackage{float}
\usepackage{graphicx, epstopdf}
\usepackage{color}
\usepackage{setspace}
\usepackage{dsfont}
\usepackage{xr}
\newcommand{\cF}{\mathcal{F}} 
\newcommand{\Mc}{\mu} 
\newcommand{\qb}{\mathbf{q}} 
\newcommand{\rb}{\mathbf{r}} 
\newcommand*{\Scale}[2][4]{\scalebox{#1}{$#2$}}

\begin{document}

\title{Chrono CDI: Coherent diffractive imaging of time-evolving samples}

\author{A. Ulvestad}
\affiliation{Materials Science Division, Argonne National Laboratory, Lemont, IL 60439, USA}
\author{A. Tripathi}
\affiliation{Mathematics and Computer Science Division, Argonne National 
Laboratory, Lemont, IL 60439, USA}
\author{S. O. Hruszkewycz}
\affiliation{Materials Science Division, Argonne National Laboratory, Lemont, IL 60439, USA}
\author{W. Cha}
\affiliation{Materials Science Division, Argonne National Laboratory, Lemont, IL 60439, USA}
\author{S. M. Wild}
\affiliation{Mathematics and Computer Science Division, Argonne National 
Laboratory, Lemont, IL 60439, USA}
\author{G. B. Stephenson}
\affiliation{Materials Science Division, Argonne National Laboratory, Lemont, IL 60439, USA}
\author{P. H. Fuoss}
\affiliation{Materials Science Division, Argonne National Laboratory, Lemont, IL 60439, USA}

\begin{abstract}
Bragg coherent x-ray diffractive imaging is a powerful technique for
investigating dynamic nanoscale processes in nanoparticles immersed in 
reactive, realistic
environments. Its temporal resolution is limited, however, by the oversampling 
requirements of 3D
phase retrieval. Here we show that incorporating
the entire measurement time series, which is typically a continuous physical 
process, into phase retrieval allows the oversampling requirement at each time 
step to be reduced leading to a subsequent improvement in
 the temporal resolution by a factor of 2-20 
times. The increased time resolution will allow imaging of faster dynamics and 
of radiation-dose-sensitive samples. This approach, which we call ``chrono 
CDI," may find use in improving time resolution in
other imaging techniques. 
\end{abstract}

\maketitle

\section{Introduction}
\label{sec:intro}
Understanding nanoscale processes 
is key to improving the performance of advanced technologies, such as  
batteries, catalysts, and fuel cells. However,
many processes occur inside devices at short length and time scales in reactive 
environments 
and represent a significant
imaging challenge. Bragg coherent diffractive imaging (BCDI) has emerged as a 
powerful 
technique for revealing
3D nanoscale structural information \cite{Pfeifer2006,Robinson2009}.
With current BCDI methods, 3D image reconstructions of nanoscale crystals have 
been used to identify and track
dislocations \cite{Ulvestad2015a,Clark2015a}, image cathode lattice strain
during battery operation \cite{Ulvestad2014a,Ulvestad2015}, indicate the 
presence of 
surface adsorbates \cite{Cha2013,Watari2011b}, and reveal twin domains 
\cite{Ulvestad2015b,Huang2015}.
The temporal resolution of current BCDI experiments, however, is
limited by the oversampling requirements for current
phase retrieval algorithms. The insight developed in this work is that, for 
most physical processes, structural evolution is a continuous process that 
introduces structural redundancy when measured as a time series. Here, we 
exploit this redundancy to allow for reduced oversampling (less than
the conventionally required factor of 2), thereby 
improving the measurement rate. In principle, this method can also be used to 
increase the spatial resolution. 

Our new approach, which we call ``chrono CDI,'' improves the temporal
resolution of BCDI by 
reducing the oversampling requirement along one dimension ($q_z$) at a given time step 
without significantly compromising image fidelity. 
To enable this capability, we designed a reconstruction algorithm that
simultaneously reconstructs all time states in a series of Bragg rocking curves by 
utilizing constraints from neighboring time steps.  In this work, the initial 
and final states are assumed to be known in real
space. In practice, this situation is achieved provided both states are measured with the 
required oversampling (OS). The time frame over which the sample is assumed to 
be static (the measurement time of an individual rocking curve) is reduced in 
chrono CDI, providing access to faster dynamics in 3D crystals. In addition,
this approach can be used to limit the radiation dose in radiation-sensitive 
samples and/or increase the spatial resolution by allowing for reduced sampling
in $q_x$ and $q_y$. 

\begin{figure}
\includegraphics[]{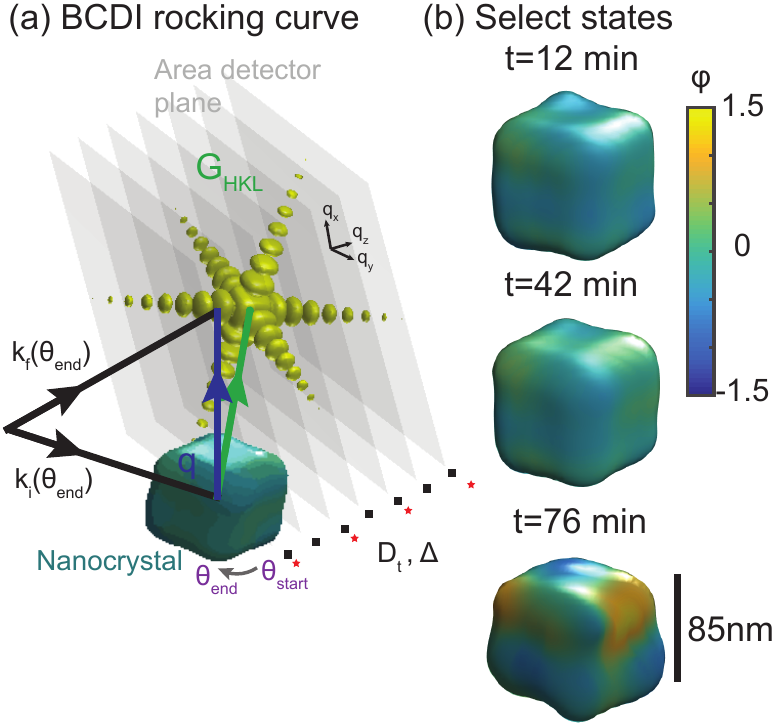}
    \caption{
     \label{fig1}
     Bragg CDI experiment of a single time-evolving nanocrystal: \textbf{(a)}
Schematic of a Bragg rocking curve. The rocking curve involves rotating the sample with respect to the 
incident x-rays $\mathbf{k}_i$. Here, $\mathbf{k}_f$ represents the scattered x-rays, 
$\mathbf{q}$ the particular scattering vector, and $\mathbf{G}_{hkl}$ the particular 
Bragg peak. The rocking curve sweeps the 2D area detector through the 3D 
volume in the $q_z$ direction. Conventionally, all 2D measurements shown by black boxes are 
required; in order to improve the time
resolution, 
only some 2D measurements (red stars) could be taken. $D_t$ denotes the full 3D diffraction
measurement and
$\Delta$ the time required to make the full measurement. \textbf{(b)}  Select 
time states from the time sequence that was reconstructed from experimental BCDI data measured 
with greater than the required oversampling. The
isosurface is drawn at a constant Bragg electron density and represents the 
shape of the 85 nm Pd nanocube.  The isosurface color
is the imaginary part of the image, the phase $\phi$,  which is proportional to 
the $u_{111}$ displacement field.}
\end{figure}

Figure~\ref{fig1}a shows a schematic of a Bragg rocking curve. The rocking curve entails 
rotating the sample 
with respect to the incident x-ray beam $\mathbf{k}_i$. 
The sample rotation, labeled schematically by $\theta_{\text{start}}$ and 
$\theta_{\text{end}}$,  displaces the scattering vector 
$\mathbf{q}=\mathbf{k}_f-\mathbf{k}_i$  
from the reciprocal-space lattice point 
$\mathbf{G}_{hkl}$, the Bragg reflection condition for the HKL lattice planes,  
so that the 3D intensity distribution (yellow isosurface) can 
be appropriately sampled and the structure of the nanocrystal (green cube) can 
be reconstructed.  The series of 2D measurements (grey planes in 
Fig.~\ref{fig1}a) are stacked to form a 3D dataset $D_t$, where $t$ represents 
a 
time index in a series of sequential rocking curve measurements. The total time for 
the measurement is $\Delta$.
Current phase retrieval algorithms require an oversampling of at least 2 of the diffraction 
pattern in all three dimensions. We refer to an oversampling of 2 as the required oversampling. 
In addition, the nanocrystal must be 
approximately static over the measurement time $\Delta$ while $D_t$ is collected, which limits the 
dynamic timescale that can be observed \cite{Xiong2014,Clark2013}. 

Figure~\ref{fig1}b shows three select time states from a time series 
during which a single crystal palladium nanocube (85 nm side length) is exposed 
to hydrogen gas. Experimental details are given in a recent publication  
\cite{Ulvestad2015NC}. The absolute value of the image (shown as an isosurface) 
corresponds to the Bragg-diffracting electron 
density \cite{Ulvestad2015b}, while the phase, $\phi$, of the image (color 
projected onto the isosurface) is proportional to a component of the vector 
displacement field $\mathbf{u}$ via $\phi=\mathbf{u}\cdot\mathbf{Q}$ 
\cite{Pfeifer2006,Chapman2006a,Newton2010}.
In this case, the Pd (111) Bragg peak was measured and $\phi\sim u_{111}$.
In the Pd nanocube, hydrogen intercalation initially causes displacement field 
changes ($t=42$ minutes) before morphological changes occur
($t=76$ minutes) due to the hydriding phase transformation 
\cite{Wicke1996,Flanagan1991}.  The time evolution of the nanocube structure 
shown in Fig.~\ref{fig1}b was determined
from BCDI experiments performed with an oversampling of 3 in $q_z$ at Sector 
34-ID-C of the Advanced Photon Source
at Argonne National Laboratory (see Experimental Details in Supplemental 
Material and Ulvestad, et al.\cite{Ulvestad2015NC}). Each complete measurement took 
approximately 2 minutes ($\Delta$ in Fig.~\ref{fig1}). 


\section{Algorithmic Approach}
\label{sec:alg}
To incorporate the redundancy in correlated time series such as those in Fig.~\ref{fig1}, 
we modify 
conventional BCDI phase retrieval algorithms. The function minimized by the error reduction 
phase retrieval algorithms is the modulus error, 
$\varepsilon_{\Scale[0.5]{\mathcal{M}}}^2$, which measures the 
agreement between the reconstruction's Fourier moduli and the measured moduli,
$$\varepsilon_{\Scale[0.5]{\mathcal{M}}}^2(\rho,D) 
=
\sum_{\qb} \Big \vert |\tilde{\rho}| - \sqrt{D}  \Big\vert^2,
\label{eq:moderror}
$$
where $\rho$ is the 3D reconstructed object (in real space), $D$ is the 3D 
far-field intensity measurement, $\qb$ is the reciprocal-space coordinate,
$\tilde{\rho}=\cF \left[ \rho \right]$, and $\cF$ is the Fourier transform. Different
choices of the function to be minimized lead to different phase retrieval algorithms 
\cite{Marchesini2007a}.  In chrono CDI,  we
include a term that depends on reconstructions at other time states,
\begin{equation}
\sum_t \bigg[
\varepsilon_{\Scale[0.5]{\mathcal{M}}}^2(\rho_t, D_t) +
\sum_{t'\neq t}
w(t,t')
\Mc(\rho_t,\rho_{t'}) 
\bigg].
\label{eq:genobjective}
\end{equation}
In this expression, $t$ indexes the time states, $w(t,t')$ is
the weight, $t\neq t'$, and $\mu(\rho_t,\rho_{t'})$ is the miscorrelation term. 
In this paper, we consider nearest-neighbor correlations in time, a
scalar weight parameter $w\geq 0$, and a functional form
for the miscorrelation of
$$
\mu(\rho_t,\rho_{t'})= \sum_{\rb} \big \vert \rho_t-\rho_{t-1} \big\vert^2 
+  \sum_{\rb} \big \vert \rho_t-\rho_{t+1} \big\vert^2.
$$
The $t$th term of the objective in
Eq.~(\ref{eq:genobjective}) then becomes
\begin{equation}
 \varepsilon_{\Scale[0.5]{\mathcal{M}}}^2(\rho_t, D_t)
+
w \left( \sum_{\rb} \big \vert \rho_t-\rho_{t-1} \big\vert^2 + \sum_{\rb} \big  
\vert \rho_t-\rho_{t+1} \big\vert^2
\right).
\label{eq:objective}
\end{equation}
Although other forms are possible, this form has the advantage of
being computationally inexpensive. The iterative algorithm is derived by minimizing
Eq.~(\ref{eq:objective}) summed over $t$ (for details, see the 
Supplemental Material and refs.\cite{Marchesini2007a,Tripathi2015}). 

\section{Numerical Results}
\label{sec:results}

To evaluate the algorithm's performance, we carry out iterative phase retrieval on 
noise-free, simulated data as well as on measured experimental data with different
 amounts of oversampling.

\subsection{Simulated Data with Required Oversampling}
\label{sec:simdata}
\begin{figure}[bt]
        \includegraphics[]{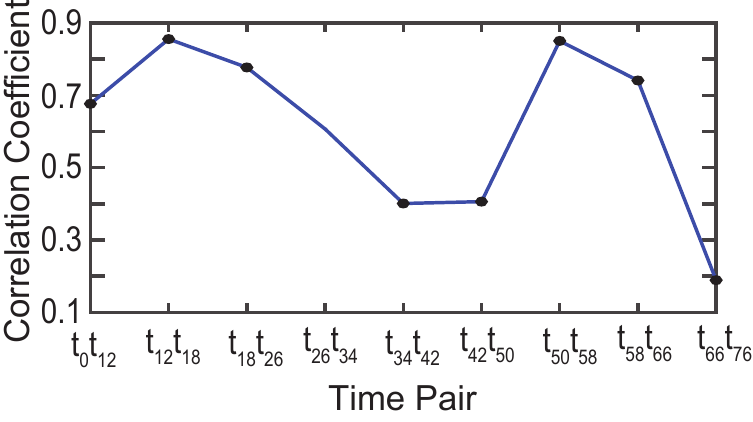}
    \caption{Nearest-neighbor correlation coefficient of $u_{111}(\mathbf{r})$ for all pairs in the 
chosen time sequence. The average nearest-neighbor correlation coefficient is 61\%.}
    \label{nncorr}
\end{figure}

In this case, the simulated data $D_t^{\rm sim}$ is generated by 
3D Fourier transforms of each complex valued Pd nanocube reconstruction in the time series 
after zero padding to meet the oversampling (OS) requirements of phase 
retrieval \cite{Livet2007}. We refer to an oversampling of 2 as the required oversampling (Req. O.S.).
 In practice, this means the cube size was half of 
the array size in all three dimensions.  
The time sequence considered consists of $t=0,12,18,26,34,42,50,58,66$, and $76$ 
minutes.  This time sequence is approximately equally spaced and, as shown in 
Fig.~\ref{nncorr}, has varying 
amounts of nearest-neighbor correlation, with an average nearest-neighbor 
correlation coefficient of 61\%. The correlation coefficient 
$c(t,t') \in [-1,1]$ is defined between two
3D displacement fields at time states $t$ and $t'$ by
\[
\footnotesize 
 \frac{ {\sum\limits_{\mathbf{r}} } \big[ u_{111}(\rb,t) -
\bar{u}_{111}(\rb,t) 
\big]  \big[ u_{111}(\rb,t') -
\bar{u}_{111}(\rb,t')
\big]}  
{\sqrt{\sum\limits_{\mathbf{r}} \big[ u_{111}(\rb,t) - \bar{u}_{111}(\rb,t)
\big]^2} \sqrt{\sum\limits_{\mathbf{r}} \big[ u_{111}(\rb,t') - \bar{u}_{111}(\rb,t')
\big]^2}} \nonumber
\]
where $u_{111}$ is the displacement field projection and 
$\bar{u}_{111}$ is the average displacement field over the particle.

\begin{figure}
\includegraphics[]{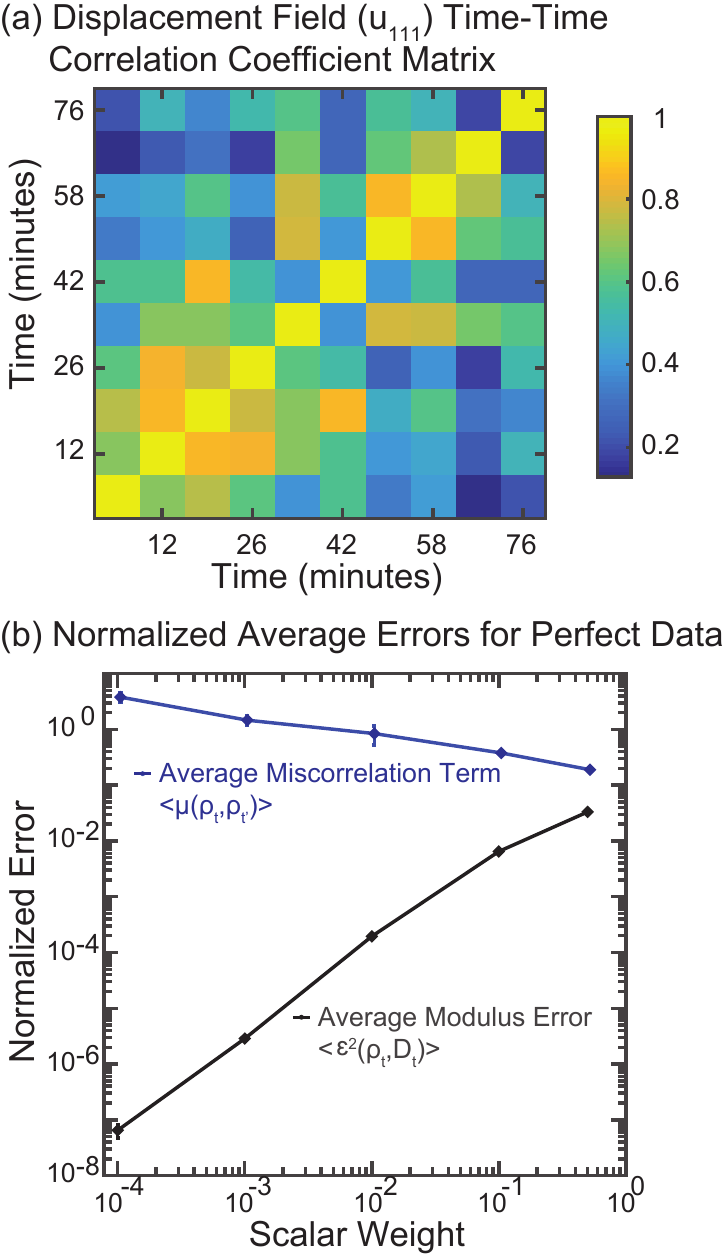}
    \caption{\textbf{(a)} Time correlation in the [111] displacement field 
projection, $u_{111}(\rb)$, during the reconstructed time sequence.
\textbf{(b)}, Average (over all time states and random starts) modulus error  
$\varepsilon_{\Scale[0.5]{\mathcal{M}}}^2 (\rho_t, D^{\rm sim}_t)$ (black) 
and 
average (over all time states and random starts) miscorrelation term $\Mc(\rho_t,\rho_{t'})$ (blue) normalized by the 
total intensity in the image as a function of the scalar 
weight. Error bars represent the standard deviation of the
average over all time states obtained from 10 different random starts.}
    \label{fig2}
\end{figure}

Figure~\ref{fig2}a shows the correlation coefficient matrix for the chosen time 
sequence. Figure~\ref{nncorr} is a plot of the super-diagonal matrix values. The chosen time 
sequence is a good balance between 
having smooth evolution between nearest neighbors and having a large change over 
the whole time sequence (both the displacement and the amplitude change 
significantly). The first numerical test of chrono CDI reconstructs the sequence 
$\rho_t$ from the sequence of $D_t^{\rm sim}$ with the required oversampling. 

The algorithm uses with random initial starts and 
alternates between the error reduction (ER) and the hybrid input-output (HIO) algorithms 
using a 
feedback parameter of $\beta=0.7$ \cite{Fienup1986}; the support 
is fixed to the size of the object and is not evolved during the iterative process. 
At iteration numbers $N=100n$, for $n=1, 2, \ldots, 18$,
 the algorithm tests whether all reconstructions are correctly oriented with
respect to the known initial ($t=0$ minutes) 
and final ($t=76$ minutes)  states by testing whether they 
are 
conjugated and reflected (``twin") solutions \cite{Fienup1986}. Although reconstructing the
``twin" image does not affect $\varepsilon_{\Scale[0.5]{\mathcal{M}}}^2$, it 
will negatively impact $\Mc$,  
resulting in an artificially high total objective. One constraint used in 
the present work is that the $\rho_t$ of the initial and final states are known 
in real and diffraction space. This constraint can be achieved by measuring diffraction datasets 
at the required oversampling before the experimental dynamics 
start and after no significant changes are seen in the diffraction data. 

Figure~\ref{fig2}b shows the errors $\varepsilon_{\Scale[0.5]{\mathcal{M}}}^2 
(\rho_t, 
D^{\rm sim}_t)$ and $\Mc(\rho_t,\rho_{t'})$, averaged over all reconstructed 
time states and over 10 random starts, as
a function of the scalar weight $w$. Both errors are normalized by the total 
intensity in the image. Figure~\ref{sampmod} in the Supplemental Material shows the modulus 
error $\varepsilon_{\Scale[0.5]{\mathcal{M}}}^2$ as a 
function of iteration number for two scalar weight values.
The initial and final states ($t=0$ and $t=76$ minutes, respectively) are known 
in real space.
When $w=0$, the modulus error is the lowest, and the miscorrelation term is the largest.
These results are expected because the data is noise-free and oversampled at the required
oversampling such 
that a unique solution is fully determined for each $D_t^{\rm sim}$. The 
weight $w=0$ 
corresponds to the case when no correlations are taken into account. 
As $w$ increases,
 $\Mc(\rho_t,\rho_{t'})$ decreases, and 
$\varepsilon_{\Scale[0.5]{\mathcal{M}}}^2$ increases for the solution set $\{\rho_t\}$ 
because their relative 
contributions to the total objective change. 
With data sampled at the required oversampling, including information from neighboring time steps in a time 
series will not improve each individual reconstruction because the complete 3D 
structural description of the sample at each time is uniquely encoded in the 3D coherent intensity pattern. 

\subsection{Simulated Data with Reduced Oversampling}
\label{sec:undersimdata}
We now explore how the additional redundancy from the time
series can compensate for reduced oversampling during the rocking curve (e.g. oversampling at less than a factor of 2)
 at a given time step
by reconstructing the time series $D_t^{\rm sim}$ discussed previously but with 
different degrees of reduced sampling in $q_z$. 
To start, every third 2D diffraction measurement of the original 84 2D  
diffraction measurements was selected to form $D_t^{\rm sim}$ for all times except
the initial and final time. This leads to data 
that has 1/3 of the required oversampling.
If such a time series were measured experimentally, the measurement time would 
be reduced by a factor of 3.
In assessing algorithm performance, the modulus error was calculated by comparing the 
far-field exit wave of the reconstructions with the data sampled at the required oversampling. 
As before, the initial and 
final states are known, the support is known, and alternating
ER/HIO is used as described previously.

\begin{figure}
        \includegraphics[]{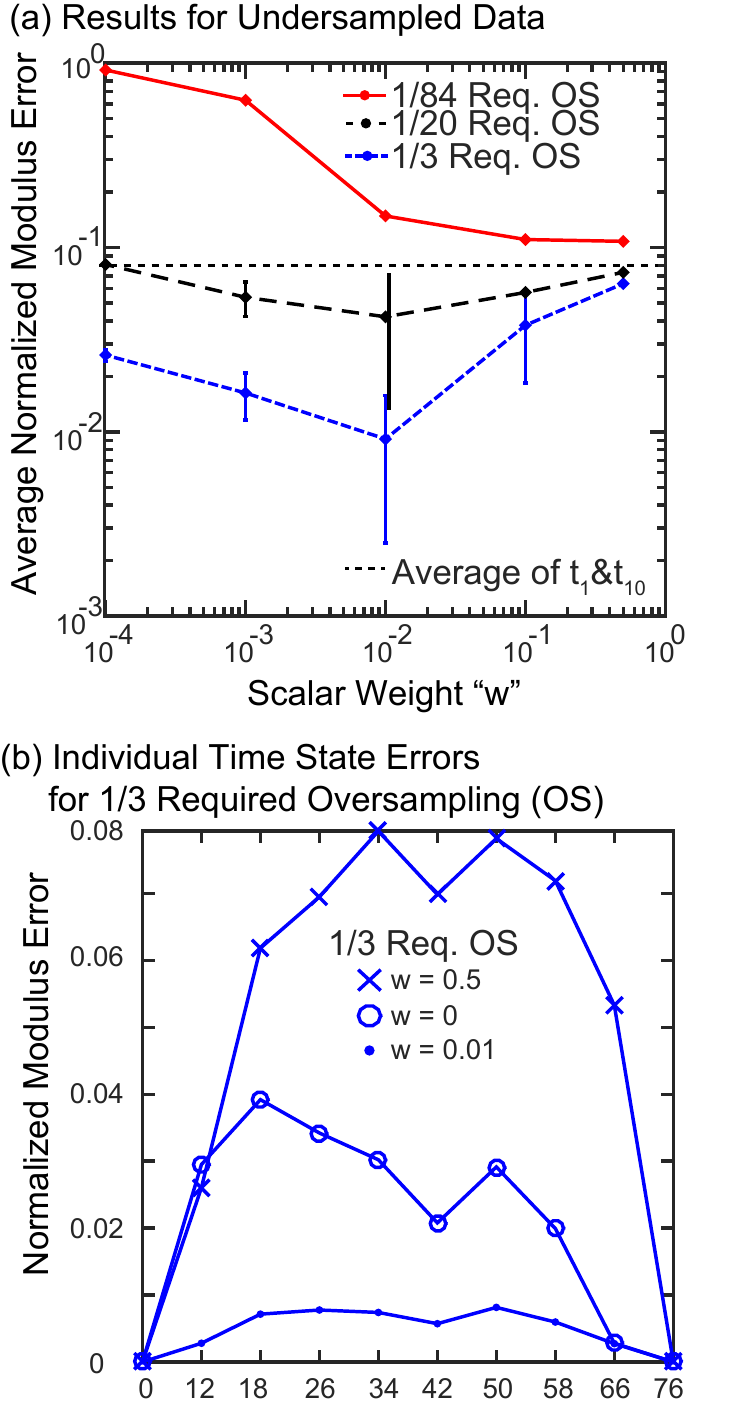}
    \caption{Results for reduced oversampling (OS). \textbf{(a)} 
The average (over all time states and random starts) normalized modulus
error as a function of the scalar weight $w$ 
for 1/3 (blue), 1/20 (black), and 1/84 (red) of the required oversampling. $w=0$ (not shown) produces the same 
average normalized modulus error values as $w=10^{-4}$.
The modulus error reported here is computed with respect to the
data with the required oversampling. Error bars represent the standard deviation of the different
average (over all time states) values obtained from 10 different random starts. Error bar for $w=0.01$
and 1/20th (black) of the required oversampling is offset for clarity. The black dashed horizontal line shows the 
normalized average modulus error using the average of the initial and final time 
states for every time state in the sequence.  \textbf{(b)} The individual time
state errors for 1/3 (blue) required oversampling for a particular random start for 
$w=0$ (open circle), $w=0.01$ (points), and $w=0.5$ (x).}
    \label{fig3}
\end{figure}

Figure~\ref{fig3}a shows the average (over all time states and random starts) modulus error 
as a function of the scalar weight $w$ 
for varying amounts of oversampling. The average normalized modulus error does not change from $w=0$ to $w=10^{-4}$. 
Unlike the results using the required oversampling (shown in 
Fig.~\ref{fig2}b) where the lowest modulus error
occurs for $w=0$ (no time correlation), in the cases where $q_z$ has been sampled at 1/3 (blue) and 1/20 (black) of the 
required oversampling, a minimum in the modulus error is observed at a value of 
$w=0.01$. 
This modulus error is computed with respect to the datasets that are 
sampled at the required oversampling, and thus the reconstructions at these minima are the 
``best" solutions. 
When only 1/84 (red) of the required oversampling is used (i.e., a single slice 
from the rocking curve), the modulus error decreases with increasing $w$ and 
approaches a constant value, which is near the normalized average modulus error 
when all the reconstructed time states are set to the average of the 
initial and final state (black dashed horizontal line). In this case, simply using an 
average of the initial and final states outperforms the reconstruction 
algorithm, indicating that there are insufficient reciprocal space constraints 
and that the oversampling in $q_z$ is too low. 

On average $w=0.01$ improves the reconstructed time sequence 
relative to $w=0$, which takes no time correlation into account, for up to 
1/20 
of the required oversampling. However, it is not clear from the plot of the 
average whether all time states are being improved equally. Figure~\ref{fig3}b shows 
the normalized modulus error at each time state for $w=0, 0.01, 0.5$ at 1/3 of 
the required oversampling for a particular random start. 
The states nearest to the known states ($t=0$ and $t=76$ 
minutes) have 
lower modulus errors, as expected. By comparing $w=0$ with $w=0.01$, we see 
that the improvement occurs in all the intermediate states except 
the state at $t=66$ minutes, 
which remains essentially unchanged. These results demonstrate that the algorithm 
improves all reconstructions, even those least correlated with their neighbors (see Fig.~\ref{nncorr} for a plot of the 
nearest neighbor correlation). 
The benefits of 
chrono CDI are clear when the required oversampling is reduced by up to a 
factor 
of approximately 1/20. In these cases, enforcing a degree of nearest-time-step, 
real-space correlation provides an additional constraint that improves the 
reconstruction at all intermediate times relative to what can be achieved using 
conventional phase retrieval. 

\begin{figure*}[tb]
        \includegraphics[]{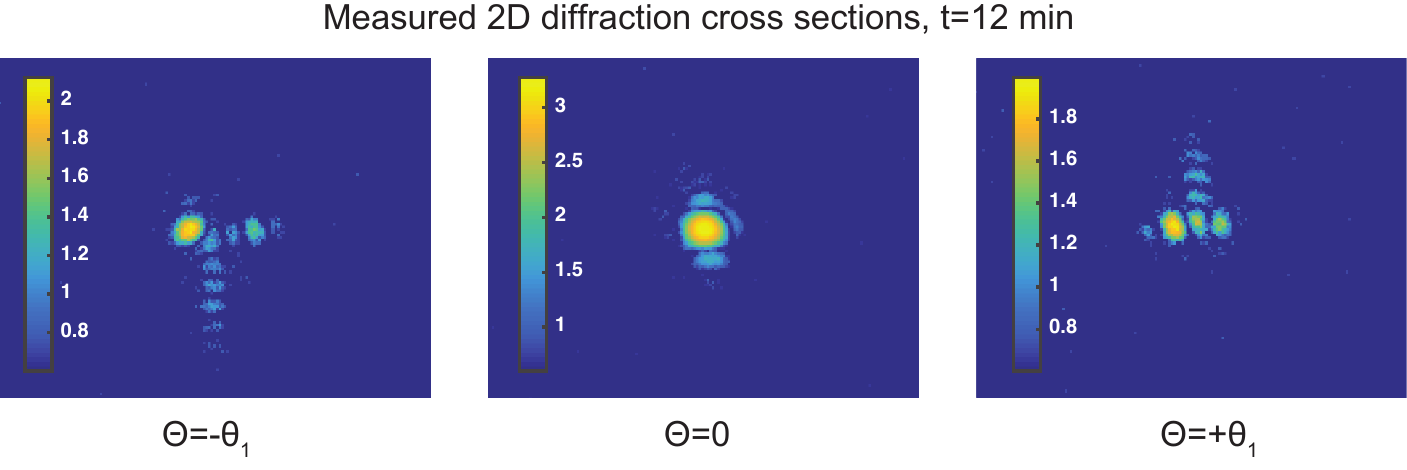}
    \caption{Three cross sections of the real data from $t=12$ minutes used to test the chrono CDI 
algorithm. The color bar is the log$_{10}$ of the number of photons. The data 
is 
oversampled by a factor of 3 in $q_z$ and has both noise and a 
finite scattering extent. These measured datasets, after background subtraction 
(1--2 
photons) and the removal of 
a number of 2D slices, are used to test the chrono CDI algorithm. Please see 
Ulvestad, et al. \cite{Ulvestad2015NC} for further details.
}
    \label{realdifdat}
\end{figure*}

\subsection{Experimental Data}
\label{sec:expdata}
We now demonstrate chrono CDI on experimental rocking curve data. To simulate
varying degrees of reduced oversampling, a subset of the original 2D measurements was selected from the 
experimental datasets. 
Figure~\ref{realdifdat} shows example 2D experimental diffraction measurements 
from the Pd (111) Bragg rocking 
curve; see Ulvestad, et al. 
\cite{Ulvestad2015NC} for more details.
%
%
 An 
oversampling of approximately 3 in $q_z$
was used during the original measurement. The 
reconstruction algorithm is the same as described previously except that the 
support is not known a priori. Instead, an initial box half the 
array size in each dimension is used, and the support is updated with the 
shrinkwrap algorithm \cite{Marchesini2003,Chapman2006a} using a Gaussian 
function with a threshold of $0.01$ and standard deviation of 1. 

\begin{figure}[t!]
        \includegraphics{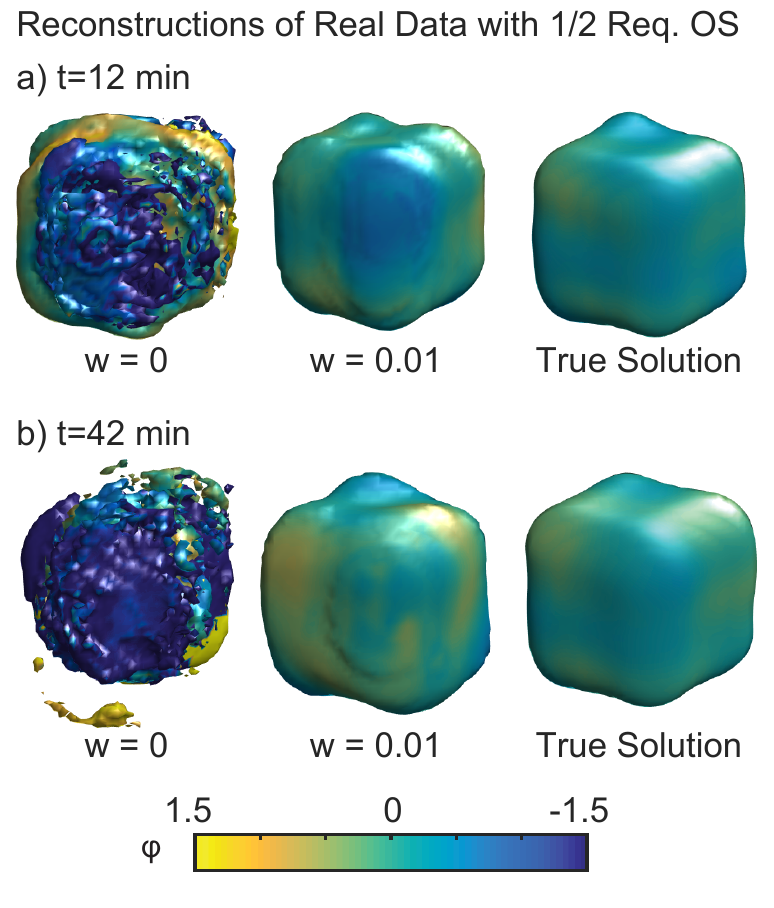}
    \caption{Reconstructions of experimental measurement data with reduced oversampling. 
 The real part of the image (shown as an 
isosurface) corresponds to the reconstructed Bragg electron density, while the 
complex part of the image (colormap projected onto the isosurface) corresponds 
to the reconstructed displacement field projection. \textbf{(a)} The $t=12$ 
minutes 
reconstructions for 1/2 of the required oversampling
for $w=0$ and $w=0.01$, and the true solution. 
\textbf{(b)} The same as \textbf{(a)} but for the $t=42$ minutes 
reconstruction.}
    \label{realdat}
\end{figure}

Figure~\ref{realdat} shows chrono CDI reconstructions for two representative 
time states of experimental diffraction data from Pd nanocubes undergoing 
structural transformations when exposed to
hydrogen gas.
As before, different amounts of oversampling were investigated. The isosurfaces 
shown correspond to the 
reconstructed Bragg electron density, while the color map  corresponds to the 
image phase $\phi$, which is proportional to the $u_{111}$ 
displacement field. Figure~\ref{realdat}a shows reconstructions when 1/2 of the required 
oversampling in $q_z$ is used. Every third slice of the original data (oversampled at a factor of 3 in $q_z$) was used to generate
data. This corresponds to an oversampling of 1, which is 1/2 the required oversampling of 2. The
reconstruction for $w=0.01$ is much improved compared with $w=0$ and is similar 
in morphology and lattice displacement to the true solution. 
Figure~\ref{realdat}b shows that the same
conclusion holds for $t=42$ minutes. See 
Figure~\ref{cxrealdat} in the Supplemental Material for central cross-sections that show the 
amplitude and phase distributions inside the crystal. 
The average normalized modulus error of the time sequence is improved from 
$0.2$ to $0.1$ by including nearest-neighbor information (via $w=0.01$). 
Although the reconstructions do not match exactly, the results convey the same 
overall physical changes in the crystal. 

\begin{figure}[tb]
        \includegraphics[width=\linewidth]{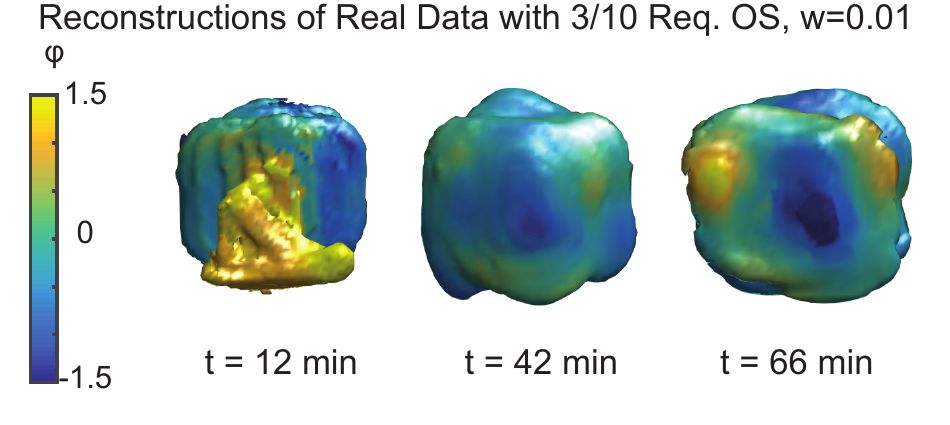}
    \caption{Reconstructions of real measurement data with 3/10 of the required 
oversampling and $w=0.01$. 
 The real part of the image (shown as an 
isosurface) corresponds to the reconstructed Bragg electron density, while the 
complex part of the image, $\phi$, is proportional to the displacement field 
projection. Three states from the reconstructed time sequence are shown. 
}
    \label{15datremsupp}
\end{figure}

Figure~\ref{15datremsupp} shows that when 3/10 of 
the required oversampling is used (corresponding to every 5th slice of the original data), 
major differences in both the reconstructed 
Bragg electron 
density and displacement fields arise as compared with the full-rocking-curve 
reconstructions. There are also disagreements in the reconstructed phases 
(proportional to the displacements). 
We therefore conclude that chrono CDI applied to this particular set of measurement data for the chosen
time sequence
could have decreased the measurement time by a factor of 2 without losing the 
essential physics of the transforming crystal. For simulated data it could have 
reduced the time by up to a factor of 20. The discrepancy is likely due to the 
finite extent in reciprocal space of the real data, noise in the data, and the 
unknown support that must be determined via the shrinkwrap algorithm during the 
reconstruction.

\section{Discussion}
\label{sec:discussion}
In this work, we have shown that our new algorithm improves the
time resolution of BCDI by a factor of 2 for experimental data and 20 for simulated data. The algorithm 
thereby 
enables BCDI 
investigations of dynamic structural processes in crystals that were previously 
out of reach and limiting radiation dose in sensitive samples.
The time resolution improvements we demonstrate are achieved by reducing the 
number of 2D measurements made 
during a 3D Bragg rocking curve, leading to datasets with less than the required
oversampling in $q_z$ at each intermediate time 
step.
The rocking curves across the entire time series are reconstructed 
simultaneously, enforcing a degree of real-space correlation between solutions 
at neighboring time steps to account for the reduced oversampling of each 
individual measurement. The algorithm and 
its variations should be useful for improving the time resolution 
of other imaging 
techniques such as ptychography and  
tomography\cite{Mohan2015,Gibbs2015,ptycho1,Hruszkewycz2013,Tripathi2011} 
where there is a continuous relationship in real space between nearest neighbor 
time states. 

\section*{Acknowledgments}
This material was based upon work supported by the U.S. Department of Energy, Office of
Science,
under Contract No. DE-AC02-06CH11357. This research used resources of the Advanced Photon Source, which is a DOE Office of Science User Facility. Work at the Advanced Photon Source was supported by the
Office of Basic Energy Sciences (BES). P.\ H.\ F., S.\ O.\ H., and G.\ 
B.\ S.\ were supported by
Basic Energy Sciences, Division of Materials Sciences and Engineering. A.\ T.\ 
and S.\ M.\ W.\ were
supported by the Office of Advanced Scientific Computing Research and the 
ROMPR project. A.\ U.\ was 
supported by an Argonne Director's postdoctoral fellowship. 

The authors thank Jesse N.\ Clark for insightful discussions.

\end{document}


\title{Chrono CDI: Coherent diffractive imaging of time-evolving samples - Supplemental Material}

\author{A. Ulvestad}
\affiliation{Materials Science Division, Argonne National Laboratory, Lemont, IL 60439, USA}
\author{A. Tripathi}
\affiliation{Mathematics and Computer Science Division, Argonne National 
Laboratory, Lemont, IL 60439, USA}
\author{S. O. Hruszkewycz}
\affiliation{Materials Science Division, Argonne National Laboratory, Lemont, IL 60439, USA}
\author{W. Cha}
\affiliation{Materials Science Division, Argonne National Laboratory, Lemont, IL 60439, USA}
\author{S. M. Wild}
\affiliation{Mathematics and Computer Science Division, Argonne National 
Laboratory, Lemont, IL 60439, USA}
\author{G. B. Stephenson}
\affiliation{Materials Science Division, Argonne National Laboratory, Lemont, IL 60439, USA}
\author{P. H. Fuoss}
\affiliation{Materials Science Division, Argonne National Laboratory, Lemont, IL 60439, USA}


\maketitle

{\let\newpage\relax\maketitle}

%
%
%


Palladium nanocubes were synthesized by using a wet chemistry method and then spin 
casted onto a silicon substrate for the measurement. The average nanocube side 
length is 85 nm. A double crystal monochromator was used to select x-rays with a photon
energy of 8.919 keV with 1 eV bandwidth and longitudinal coherence length of 0.7 $\mu$m. A 
set of Kirkpatrick Baez mirrors was used to focus the beam to approximately 
1.5$\times$1.5 $\mu m^2$. The silicon substrate was placed into a gas flow cell. 
The detector was set to a (111) Pd Bragg peak angle and the sample was
scanned in the x-ray beam until a (111) Pd diffraction peak illuminated the 
detector. The particles are randomly oriented and therefore typically at 
most one diffraction pattern illuminates the area detector. 

 The rocking curve around the (111) Bragg reflection was collected by recording 
2D coherent diffraction patterns with an x-ray sensitive area detector (Medipix 
3, 55 $\mu$m pixel size) placed at 0.26 m away from the sample around an angle 
of 2$\theta = 36\deg$. An angular range of  $(\Delta \theta = \pm 0.15\deg)$ was 
sampled at 51 equally spaced intervals. The total time for each 3D measurement 
was approximately 2 minutes. Two to three full 3D datasets were taken for each 
palladium nanocrystal in a helium environment. The 3.6\% mole fraction H$_2$ gas in 
He was then flowed to increase the partial pressure of H$_2$ above zero. A Mylar 
window on the gas flow cell allows for a pressure slightly greater than 
atmospheric pressure. The pressure was left constant while 3D datasets were 
continuously collected at approximately 2-minute intervals. For further details, 
see the supplemental material of reference \cite{Ulvestad2015NC}.

\paragraph{Algorithm Derivation.}
The algorithm's objective function is defined as
\[f(\rho_1, \ldots, \rho_T) = 
\sum_{t=2}^{T-1}
\| \rho_t - \pimj \rho_t\|^2
+
w \sum_{t=1}^{T-1} \|\rho_t - \rho_{t+1}\|^2,
\]
where $\rho_t$ is the reconstruction at time $t$, $\pimj$ is the modulus projection operator,
$w$ is the scalar weight, $T$ is the total number of time states, and 
$\|A\|^2=\sum_{\mathbf{r}} |A(\mathbf{r})|^2$. 
The Wirtinger derivative with respect to one of the time states is 
\[
\frac{\partial}{\partial \rho_t}f =  \left(\rho_t - \pimj \rho_t\right)
+ w\left(2\rho_t - \rho_{t+1} - \rho_{t-1} \right).
\]

A first-order block-coordinate descent method (updating the objects in the
temporal order $\rho_2, \ldots , \rho_{T-1}$) then has an iterative update of 
the 
form
\begin{eqnarray*}
\rho_t^{(k+1)} 
&=&
\rho_t^{(k)} - \frac{\partial }{\partial \rho_t}f 
\left(\rho_1^{(k+1)}, \ldots, \rho_{t-1}^{(k+1)}, \rho_t^{(k)}, \ldots ,
\rho_T^{(k)}\right )
\\
&=& 
-2w\rho_t^{(k)} + \pimj \rho_t^{(k)}
+ w \rho_{t+1}^{(k)} + w \rho_{t-1}^{(k+1)},
\end{eqnarray*}
for $t=2, \ldots, T-1$, and $k=0, 1, \ldots, N$ for $N$ the total number of 
iterations.
 Note that $\rho_1^{(k)}$ and
$\rho_{T}^{(k)}$ are known (and constant) for all $k$. To make the algorithm 
similar to 
the standard ER, we enforce a support constraint:
\[\footnotesize 
\rho_t^{(k+1)} = \pi_{S_t} 
\left( -2w\rho_t^{(k)} + \pimj \rho_t^{(k)}
+ w \rho_{t+1}^{(k)} + w \rho_{t-1}^{(k+1)}
\right).
\]
The most recent information is always used
(hence the differences in the superscripts in the above updates), and 
the support constraint is applied for the $t$th
object (since each object may have its own support).

\begin{figure*}[tbh]
        \includegraphics[width=.9\linewidth]{sampmod.pdf}
    \caption{Normalized modulus error evolution for the eight reconstructed time 
states for noise free sufficiently oversampled data: \textbf{(a)} 
Normalized modulus error as a function of iteration number for $w=0$. 
\textbf{(b)} Evolution for $w=0.5$. }
    \label{sampmod}
\end{figure*}

Following a similar argument (and defining $\Pi_{S_t^c} = 1-\Pi_{S_t}$), we
obtain the HIO-like update
\begin{eqnarray*}
 \rho^{(k+1)}_t 
%
&=& 
(1- \alpha (1+2w)\Pi_{S_t} +2w\beta
\Pi_{S_t^c})  \rho^{(k)}_t
\\
&&
+
\left(\alpha
\Pi_{S_t}
-\beta
\Pi_{S_t^c}
\right)
\pimj \rho_t^{(k)}
\\
&&
+
w\left(
\alpha
\Pi_{S_t}
- \beta
\Pi_{S_t^c}\right)
(  \rho_{t+1}^{(k)} +  \rho_{t-1}^{(k+1)} ),
\end{eqnarray*}
for $t=2, \ldots, T-1$ and $k=0, 1, \ldots, N$. 
We used $(\alpha,\beta) = (1,0.7)$ in our reconstructions.

\begin{figure}[tbh]
        \includegraphics[]{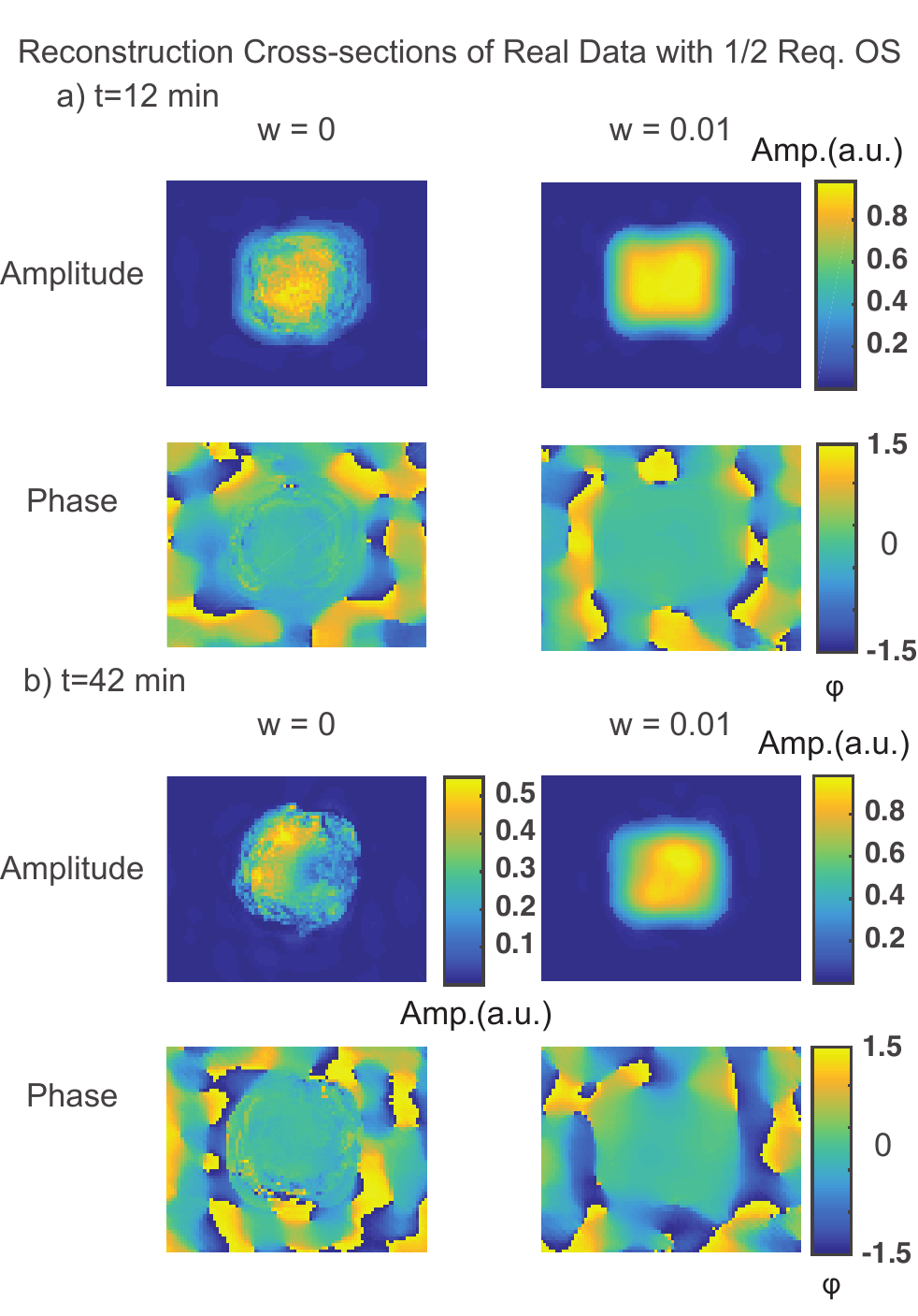}
    \caption{Cross sections of reconstructions from real data showing the 
amplitude and phase for two time states with 1/2 the required oversampling. 
 The images shown are cross sections of the states shown as isosurfaces in 
Fig.~\ref{realdat}. Shown are $w=0$ and $w=0.01$.
}
    \label{cxrealdat}
\end{figure}